\documentclass[aps,superscriptaddress]{revtex4-2}
\usepackage[colorlinks=true, pdfstartview=FitV, linkcolor=blue, citecolor=red, urlcolor=magenta, breaklinks=true]{hyperref}
\usepackage{graphicx}  
\usepackage{amsfonts}
\usepackage{color}
\usepackage{epsfig}
\usepackage{amssymb}
\usepackage{amsmath}
\usepackage{mathrsfs}
\usepackage{subfig}
\usepackage{multirow}
\usepackage{comment}
\usepackage{xcolor}
\usepackage{caption}

\begin{document}
\title{Thin-shell gravastar in a noncommutative BTZ geometry}
\author{A.~T.~N.~Silva}
\email{andersont.fisica@gmail.com}
\affiliation{Departamento de F\'isica, Universidade Federal de Sergipe,
49100-000 Aracaju, Sergipe, Brazil}
\author{M. A. Anacleto}
\email{anacleto@df.ufcg.edu.br}
\affiliation{Departamento de F\'{\i}sica, Universidade Federal de Campina Grande
Caixa Postal 10071, 58429-900 Campina Grande, Para\'{\i}ba, Brazil}
\author{L. Casarini}
\email{lcasarini@academico.ufs.br}
\affiliation{Departamento de F\'isica, Universidade Federal de Sergipe,
	49100-000 Aracaju, Sergipe, Brazil}
   
\begin{abstract} 
In this paper, we build a thin-shell gravastar model within a noncommutative BTZ geometry. 
For this, we consider a noncommutative BTZ metric in the inner region and a geometry associated with a BTZ solution in the outer region, joined by the generalized join technique. 
After investigating the inner spacetime, surface and outer spacetime, we observe that there is a surface energy density and surface pressure, such as to make gravastar stable. 
This effect persists even when the cosmological constant is zero. 
Besides, we found a bound for the noncommutativity parameter.
In addition, we examine the thermodynamics of the noncommutative BTZ black hole in Schwarzschild-type form in three-dimensional spacetime. We also check the stability condition by calculating the specific heat capacity, as well as for the formation of black hole remnants.

\end{abstract}
\maketitle

\section{Introduction}
The gravastar (gravitational vacuum star) model was initially proposed by Mazur and Mottola~\cite{Mazur:2001fv,Mazur:2004fk} and has attracted attention as a possible substitute for black holes. 
This model is of great interest because it could solve the singularity and information paradox problems. 
Thus, different approaches have been introduced to explore gravastar solutions.
Indeed, in this model, a massive star in its final stages could end up as a gravastar, a very compact object 
whose radius would be very close to the Schwarzschild radius with no event horizon or central singularity. 
To this end, phase transitions are expected to occur at or near where the event horizon originally formed~\cite{Chapline:2004jfp}. 
The interior of what would have been the black hole is replaced by a suitably chosen portion of de-Sitter spacetime with an equation of state $p = -\rho$ {(dark energy)}, surrounded by a thin layer of ultra-hard matter with $p = + \rho$~\cite{Chirenti:2007mk}, then, the noncommutative BTZ type solution is suitably combined in its exterior, which has as its main characteristic being in (2+1) dimensions.
In three-dimensional space-time, introducing the negative cosmological constant $\Lambda=-l^{-2}$~\cite{Banados:1992wn} into Einstein's equations, Barnardos, Teitelboim, and Zanelli obtained a BTZ solution for Black Hole, characterized by expressing asymptotically that its metric is close to the anti-de Sitter solution, rather than the Minkowski solution.
{In addition, the equation of state $p=\omega \rho$ with $\omega=-1$, also known as false vacuum, degenerate vacuum or $\rho$-vacuum, corresponds to a repulsive pressure. Furthermore, this repulsive pressure would be related to the accelerated expansion of the Universe with the cosmological constant $\Lambda$ acting as the dark energy responsible for inflation in the early Universe.}

Noncommutativity is another way to construct regular black holes without singularities with minimum length $\sqrt{\theta}$.
Nicolini \textit{et al.}~\cite{Nicolini:2005vd} implemented noncommutativity in black holes by considering mass density as a Gaussian distribution with minimum length $ \sqrt{\theta} $.
Since then, several works in black hole physics inspired by noncommutative geometry are found in the literature — see~\cite{Nicolini:2008aj,Szabo:2006wx} for comprehensive reviews. 
In particular, Lobo and Garattini~\cite{Lobo:2010uc} analyzed gravastar solutions in noncommutative geometry and studied their physical characteristics. 
In~\cite{Ovgun:2017jzt}, \"Ovg\"un \textit{et al.} introduced a thin-shell charged gravastar model in four dimensions in noncommutative geometry, and the stability of such a gravastar has been investigated.
{In the context of the modified theory of gravity, some work on gravastar has been carried out. 
In~\cite{Yousaf:2019zcb} the authors investigated the effects of electromagnetic field on the isotropic
spherical gravastar models in metric $f(R, T)$ gravity. 
In~\cite{Shamir:2018qhq}  gravastars solutions in gravity $f(\mathcal{G},T)$ have been examined. 
Gravastars in $f(R, G)$ gravity was explored in~\cite{Shamir:2020apc}.
M. Farasat Shamir analyzed the solutions of the Einstein-Maxwell field equations for the study of compact stars~\cite{Shamir:2020gzh}. In~\cite{Shamir:2020ckh} the physical characteristics of compact stars by exploiting the so-called Noether symmetry approach have been studied.}
Alternatively, we can introduce noncommutativity into black holes through a Lorentzian distribution~\cite{Nozari:2008rc}. 
By applying the study of Lorentzian distribution on the thermodynamics of black holes~\cite{Liang:2012vx,Anacleto:2020efy,Anacleto:2020zfh}, scattering processes~\cite{Anacleto:2019tdj,Anacleto:2022shk,Jha:2022nzd}, quasinormal modes (and shadow black holes)~\cite{Campos:2021sff,Jha:2022bpv,Zeng:2021dlj,Saleem:2023pyx} and holographic Einstein rings~\cite{Hu:2023eow} have been explored. 
Recently, in~\cite{Anacleto:2022sim}, the minimum length effect on the BTZ metric was introduced by considering the ground state probability density of the hydrogen atom, and the thermodynamic properties were examined. 
However, studies of gravastars in lower dimensions in noncommutative geometry have been little explored. 
Thus, we will focus on the gravastar model in the thin-shell noncommutative BTZ black hole by adopting a Lorentzian smeared mass distribution.

A black hole with dimension (2+1) makes a good and relatively simple laboratory to explore and test some of the ideas behind the AdS/CFT~\cite{Maldacena:1997re} correspondence. In addition to these reasons, the study of the thermal properties of three-dimensional black holes has drawn attention~\cite{Ashtekar:2002qc}, as well as the analysis of general aspects of the physics of black holes at the quantum level. 
The study of gravity in (2+1)-dimensions can help to understand fundamental aspects of gravity at the classical and quantum levels and new insights into gravity in (3+1)-dimensions. 
In~\cite{Rahaman:2011we}, Rahaman \textit{et al.} implemented a spherically symmetric neutral model of gravastar in (2 + 1) anti-de Sitter spacetime. 
Later the authors in~\cite{Rahaman:2012wc} analyzed the charged gravastar solution in (2+1)-dimensions, showing that the model is non-singular. 
In ~\cite{Banerjee:2016bry}, the authors have investigated the stability of gravastar in lower dimensions in noncommutative geometry. 
The stability of three-dimensional AdS gravastar has also been explored in the context of rainbow gravity~\cite{Barzegar:2023ueo}.
Also, other types of similar objects were considered in~\cite{Khlopov:1985jw,Konoplich:1999,Khlopov:2000,Khlopov:2008qy,Dymnikova:2015yma}.

In this work, we will investigate a type of gravastar in which we will consider a noncommutative BTZ metric in the inner region and a geometry associated with a BTZ solution in the outer region, both united, at their limits, by a thin shell. Thus, we will verify the energy conditions, based on its surface energy density, $\sigma$, and surface pressure, $\mathscr{P}$. 
As a result we show that the conditions of null and strong energy are satisfied even with the null cosmological constant.
Initially, we perform thermodynamic analysis of the noncommutative BTZ black hole in Schwarzschild-type form in three-dimensional spacetime.

The paper is organized as follows. In Sec.~\ref{secbtznc}, we introduce a noncommutative BTZ metric by considering a Lorentzian mass distribution, and we analyze the effect of noncommutativity in the calculation of the Hawking temperature, entropy and the specific heat capacity. 
In Sec.~\ref{stbtzgrav} we present the structural equations of gravastar, we examine the
matching conditions at the junction interface, and we find the surface energy density and the surface pressure.
In Sec.~\ref{Conc} we make our final considerations.

\section{BTZ metric on a noncommutative geometry}
\label{secbtznc}
In this section, we construct the BTZ metric in the noncommutative background, and 
then, we incorporate noncommutativity through a Lorentzian mass density of a spherical region of radius $r$, given by~\cite{Liang:2012vx,Anacleto:2020efy}
\begin{eqnarray}
\rho_{\theta} = \frac{M_0 \sqrt{\theta}}{2 \pi (r^{2} + \theta)^{3/2}}, 	
\end{eqnarray} 
here $\theta$ is the noncommutative parameter with dimension of length$^2$  
and $M_0$ is the total mass spread over the entire linear sized region $\sqrt{\theta}$. In this case, the ``stained'' mass is distributed as follows~\cite{Anacleto:2020efy}:
\begin{eqnarray}
\mathcal{M} = \int_{0}^{r} \rho(r)2 \pi r dr = M_0\left(1 - \frac{\sqrt{\theta}}{\sqrt{r^{2} + \theta}} \right). 	
\end{eqnarray} 
Note that when $r \rightarrow \infty$, the noncommutative parameter disappears, thus becoming a point mass with value $M_0$, 
losing its noncommutative characteristic.
Now, using the $\mathcal{M}$ mass distribution, we have the metric of the noncommutative, non-rotating BTZ black hole which is given by~\cite{Anacleto:2020efy}:
\begin{eqnarray}
ds^{2} =  -f(r)dt^{2} + f(r)^{-1}dr^{2} + r^{2}d\phi^{2},
\end{eqnarray}
where the metric function reads
\begin{eqnarray}
f(r)=-\mathcal{M} + \frac{r^2}{l^2} = -M_0 \left( 1 - \frac{\sqrt{\theta} }{\sqrt{r^{2} + \theta}}\right) + \frac{r^2}{l^2}.
\end{eqnarray} 
{Here $l$ is the radius of curvature and that provides the length scale needed to have an event horizon radius.}

For $\theta\ll 1$, we can write the metric function as follows~\cite{Anacleto:2020efy}:
\begin{eqnarray}
f(r)= -M_0 + \frac{M_0\sqrt{\theta}}{r}  + \frac{r^2}{l^2} + \mathcal{O}(\theta^{3/2}).
\end{eqnarray}
In this approximation, a term, ${M_0\sqrt{\theta}}/{r}$, of the Schwarzschild type is generated as an effect of noncommutativity. 
The impact of this term on the thermodynamics of the BTZ black hole was investigated in~\cite{Anacleto:2020efy}, showing that a logarithmic correction term for entropy is obtained. 
Also, by calculating the specific heat capacity its stability analysis was verified.
As a result, it becomes a remnant of a black hole with a minimum mass dependent on the parameter $\theta$.

By setting $f(r)=0$, the event horizon is given by~~\cite{Anacleto:2020efy}
\begin{eqnarray}
\label{rh}
\tilde{r}_h\approx r_h - \frac{\sqrt{\theta}}{2},  \quad \mbox{or} 
\quad 
\Tilde{M}\approx \frac{r^2_h}{l^2}\left(1- \frac{\sqrt{\theta}}{r_h}\right)
= M_0\left(1- \sqrt{\frac{\theta}{l^2 M_0}}\right),
\end{eqnarray}
where $r_h=\sqrt{l^2M_0}$ is the event horizon of the usual BTZ black hole. 

For the Hawking temperature, we have~\cite{Anacleto:2020efy}
\begin{eqnarray}
\Tilde{T}_H=\frac{\Tilde{r}_h}{2\pi l^2} - \frac{M_0\sqrt{\theta}}{4\pi \Tilde{r}^2_h}.   
\end{eqnarray}
In terms of $r_h$, we obtain
\begin{eqnarray}
\Tilde{T}_H=\frac{{r}_h}{2\pi l^2} - \frac{\sqrt{\theta}}{4\pi l^2}- \frac{M_0\sqrt{\theta}}
{4\pi \left({r}_h - \sqrt{\theta}/2\right)^2}.   
\end{eqnarray}
The result can be expressed in Schwarzschild-type form as follows:
\begin{eqnarray}
\mathcal{T}_H=\frac{\Tilde{T}_H}{M_0}=\frac{1}{2\pi\left[r_h +\sqrt{\theta} + \dfrac{\theta}{2r_h} \right]}.   
\end{eqnarray}

\begin{figure}[!htb]	
\centering
\includegraphics[scale=.65]{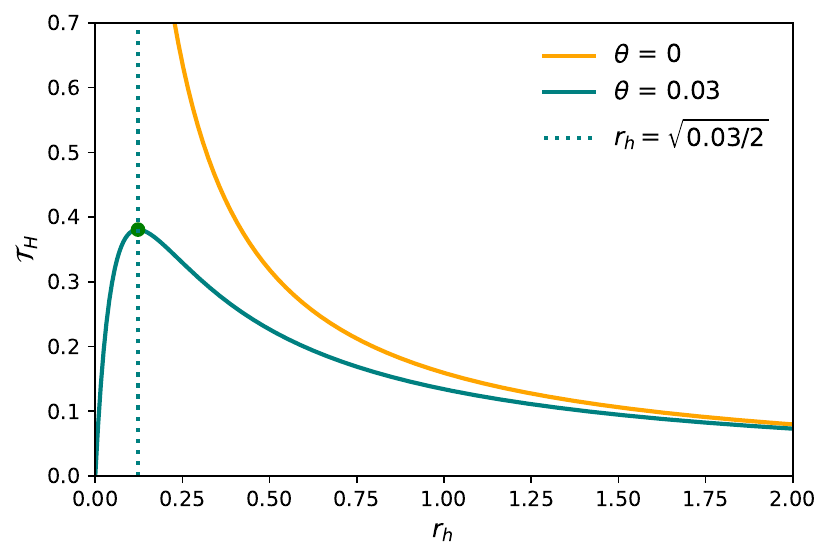}
\captionsetup{justification=raggedright,singlelinecheck=false}
\caption{\footnotesize{Hawking temperature as a function of the horizon radius $r_{h}$ for  $\theta = 0$ (orange) and $\theta = 0.03$ (teal). The green dot indicates the maximum Hawking temperature corresponding to $r_h = \sqrt{\theta/2}$ (dotted line), in the case of  $\theta = 0.03$.} { We chose this value $\theta$ because there is a noncommutativity requirement with small parameter values. However, it is possible to vary $\theta$, leading to changes in the graph: for values greater than $\theta=0.03$, the graph will be below what is shown, and for smaller values, it will be higher.}
}
\label{fg0}
\end{figure}

In Fig.~\ref{fg0}, we have Hawking temperature as a function of the horizon radius, $r_h$.  
As shown in the graph, we obtain the Hawking temperature for $\theta = 0$ and $\theta = 0.03$.

Note that the Hawking temperature reaches a maximum point before going to zero when the horizon radius, $r_h$ tends to zero, as shown in Fig.~\ref{fg0}. 
Therefore, noncommutativity has the effect of removing the Hawking temperature singularity.
Furthermore, from the condition $d\mathcal{T}_H/dr_h=0$, we obtain $r_{min}=\sqrt{\theta/2}$.

Next, for $r_h=r_{min}=\sqrt{\theta/2}$, we find that the maximum temperature is given by
\begin{eqnarray}
\mathcal{T}_{max}=\frac{\mathcal{T}_H}{2+\sqrt{2}}=\frac{1}{2\pi(2+\sqrt{2})r_{min}}=\frac{1}{2\pi(1+\sqrt{2})\sqrt{\theta}} .   
\end{eqnarray}
In particular, for $\theta=0.03$, we have $r_{min}=0.122474$ and $\mathcal{T}_{max}=0.380613$.
Now, we will determine the entropy by applying the following equation
\begin{eqnarray}
 S=\int \frac{1}{\mathcal{T}_{H}}\frac{\partial \Tilde{M}}{\partial{r}_h} d{r}_h,  
\end{eqnarray}
where, from Eq.~\ref{rh}, we have
\begin{eqnarray}
\frac{\partial \Tilde{M}}{\partial{r}_h}\approx\frac{2{r}_h}{l^2} - \frac{\sqrt{\theta}}{l^2}
=\frac{2M_0}{r_h}\left(1-\frac{\sqrt{\theta}}{2r_h} \right).  
\end{eqnarray}
So, we obtain
\begin{eqnarray}
\mathcal{S}&=&\frac{S}{M_0}=\int  \frac{2}{r_h\mathcal{T}_{H}}\left(1-\frac{\sqrt{\theta}}{2r_h} \right)d{r}_h
=4\pi\int \left[1 + \frac{\sqrt{\theta}}{2r_h}\right]dr_h,
\\
&=&4\pi r_h + 2\pi\sqrt{\theta}\ln r_h.
\end{eqnarray}
Therefore, we find a logarithmic correction term for the entropy of the noncommutative BTZ black hole. 
Moreover, for $\theta=0$, we recover the entropy of the usual BTZ black hole.

Now, we can determine the specific heat capacity through the relationship
\begin{eqnarray}
C=\frac{\partial\Tilde{M}}{\partial r_h}\left(\frac{\partial\mathcal{T}_H}{\partial r_h} \right)^{-1}.    
\end{eqnarray}   
Hence, we find
\begin{eqnarray}
\mathcal{C}=\frac{C}{M_0}=-4\pi r_h \left(1 + \frac{3}{2}\frac{\sqrt{\theta}}{r_h}\right) 
\left(1+\frac{1}{r_h}\sqrt{\frac{\theta}{2}}\right)\left(1-\frac{1}{r_h}\sqrt{\frac{\theta}{2}}\right).\\
\end{eqnarray}
By setting $r_h=\sqrt{\theta/2}$, the specific heat capacity cancels out and the BTZ black hole in Schwarzschild-type form stops evaporating completely.
Thus, becoming a black hole remnant.
Moreover, since $r_{min}=\sqrt{l^2M_{min}}$, we have the following minimum mass
\begin{eqnarray}
 M_{min}=\frac{r^2_{min}}{l^2}=\frac{\theta}{2l^2}=-\frac{\Lambda\theta}{2}.   
\end{eqnarray}
Thus, for a positive minimum mass ($M_{min}>0$) the cosmological constant must be negative ($\Lambda <0$).
In addition, for $\theta=0$, we have $\mathcal{C}=-4\pi r_h$ which is the specific heat capacity of the Schwarzschild black hole projected in 3 dimensions.

\begin{figure}[!h]	
\centering
\includegraphics[scale=.65]{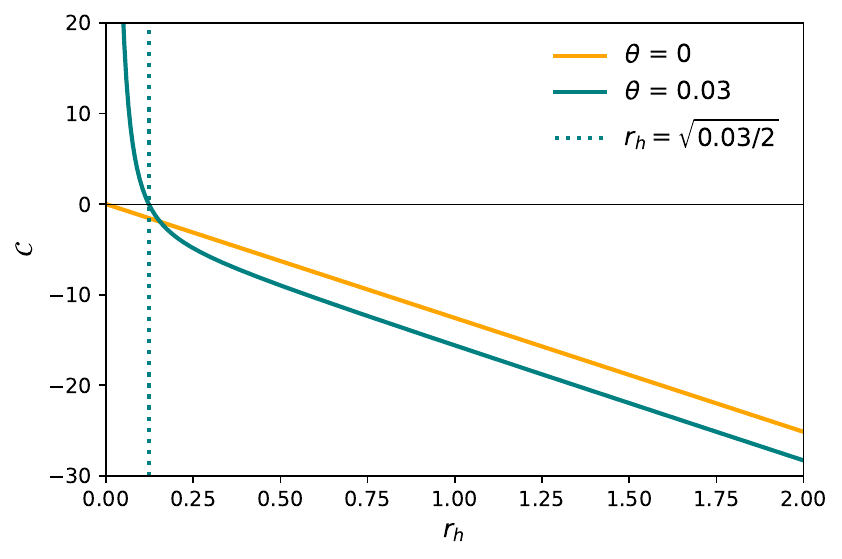}
\captionsetup{justification=raggedright,singlelinecheck=false}
\caption{\footnotesize{Specific heat capacity $\mathcal{C}$ as a function of the horizon radius $r_{h}$ for $\theta=0$ (orange) and $\theta=0.03$ (teal). 
The dotted line limits above the stability region when $\theta=0.03$. By definition this region occurs when $\mathcal{C}$ is positive, thus when $0< r_h \leq r_{min}=\sqrt{\theta/2}$.}}
\label{fg001}
\end{figure}

In Fig.~\ref{fg001}, we have specific heat capacity as a function of the horizon radius, $r_h$.  
As shown in the graph, we obtain the specific heat capacity for $\theta = 0$ and $\theta = 0.03$. 
Thus, for $0< r_h \leq r_{min}=\sqrt{\theta/2}$ reaches the black hole stability region with a positive specific heat capacity.

\section{STRUCTURE EQUATIONS OF NONCOMMUTATIVE BTZ GRAVASTARS}
\label{stbtzgrav}
To build the BTZ gravastars, we first consider two manifolds of spacetime, inspired by noncommutative geometry. The outer is defined by $M_{+}$, and the inner is $M_{-}$~\cite{Ovgun:2017jzt}. Then we join them together using the cut and paste method in a surface layer, which in this work will be called $\Sigma$~\cite{Lobo:2015lbc}. The metric of the exterior is the non-singular 
anti-de Sitter spacetime in (2 + 1)-dimensions~\cite{Banados:1992wn}:
\begin{equation}
ds^{2} = -f(r)_{+} dt_{+}^{2} + f(r)_{+}^{-1}dr_{+}^{2} + r_{+}^{2}d\phi_{+}^{2},
\end{equation}
where $t$ is the physical time in the outer region, and
\begin{eqnarray}
\label{frl}
f(r)_+=-M_0 + \frac{r^{2}}{l^{2}}=M - \Lambda r^2.  
\end{eqnarray}
Here $M=-M_0$ and $\Lambda=-1/l^2$ is the cosmological constant.

We can also write the metric function $f(r)_+$ in Schwarzschild-type form as follows:
\begin{eqnarray}
f(r)_{+} = 1- \frac{b_{+}}{r},
\end{eqnarray}
being
\begin{eqnarray}
b_+=-r\left(M - \Lambda r^2\right) +r.
\end{eqnarray}
And the metric inside is given by the noncommutative BTZ geometry. So, we have
\begin{eqnarray}
ds^{2} =  -g(r)_{-}dt_{-}^{2} + f(r)_{-}^{-1}dr_{-}^{2} + r_{-}^{2}d\phi_{-}^{2},
\end{eqnarray}
where 
\begin{eqnarray}
g(r)_{-} = M \left( 1 - \frac{\sqrt{\theta} }{\sqrt{r^{2} + \theta}}\right) - \Lambda r^2,
\label{at01}
\end{eqnarray}
and
\begin{eqnarray}
f(r)_{-} = 1- \frac{b_{-}}{r}, 
\end{eqnarray}
where
\begin{eqnarray}
b_{-}=-r\, M \left( 1 - \frac{\sqrt{\theta} }{\sqrt{r^{2} + \theta}}\right) + r.   
\end{eqnarray}
Here $\pm$ represents the outer and inner geometry, respectively.

\subsection{Transition layer}
The distributions, both inside and outside, are bounded by isometric hypersurfaces $\Sigma_{+}$ and $\Sigma_{-}$. Our goal is to join $M_{+}$ and $M_{-}$ in their limits to obtain a single variety $M$ such that $M = M_{+} \cup M_{-}$ so that, in these limits, $ \Sigma = \Sigma_{+} = \Sigma_{-}$. So, to calculate the components of the energy-momentum tensor, we will use the intrinsic metric as follows~\cite{Lobo:2015lbc}:
\begin{equation}
ds^{2}_{\Sigma} = -d\tau^{2} + a(\tau)^{2} (d\theta^{2} + \sin^{2} \theta \,\ d \phi^{2}),
\end{equation}
where $\tau$ is the physical time in the junction surface. As we are working in (2+1)-dimensions, we assume, $d \phi^{2} = 0$. 

The junction surface is given by $x^{\nu} (\tau,\theta,\phi) = (t(\tau), a(\tau), \theta)$, where the unit normal vectors with respect to this surface are the following~\cite{MartinMoruno:2011rm}:
\begin{equation}
n^{\mu}_{+} = \left(\frac{1}{M - \Lambda a^{2}}\dot{a},\sqrt{M - \Lambda a^{2} + \dot{a}^{2}}, 0 \right), 
\end{equation}
and
\begin{equation}
n^{\mu}_{-} = \left(\dfrac{1}{M \left( 1 - \dfrac{\sqrt{\theta} }{\sqrt{a^{2} + \theta}}\right) }\dot{a},\sqrt{M \left( 1 - \dfrac{\sqrt{\theta} }{\sqrt{a^{2} + \theta}}\right)  + \dot{a}^{2}}, 0 \right), 
\end{equation}
where the dot over $a$ represents a derivative with respect to $\tau$. 

The extrinsic curvatures are calculated by the following equation~\cite{Visser:1995cc}: 
\begin{equation}
K^{\psi \pm}_{\,\ \psi} = \frac{1}{a}\sqrt{1 - \frac{b_{\pm}(a)}{a} + \dot{a}^{2}}, 
\end{equation}
and					
\begin{equation}
K^{\tau\pm}_{\,\ \tau} = \left\lbrace \dfrac{ \ddot{a} + \dfrac{ b_{\pm}(a) - b^{'}_{\pm}(a)a}{2a^{2}} }{\sqrt{1 - \dfrac{b_{\pm}(a)}{a} + \dot{a}^{2}}}\right\rbrace.
\end{equation}
Thus, the extrinsic curvatures in the outer region are given by
\begin{equation}
K^{\psi +}_{\,\ \psi} = \frac{1}{a}\sqrt{M - \Lambda a^{2} + \dot{a}^{2}},  
\end{equation}
\begin{equation}
K^{\tau +}_{\,\ \tau} = \left\lbrace \frac{\ddot{a} - \Lambda {a}^2}{\sqrt{M - \Lambda a^{2} + \dot{a}^{2}}}\right\rbrace, 
\end{equation}
and in the interior region, we have
\begin{equation}
K^{\psi -}_{\,\ \psi} = \dfrac{1}{a}\sqrt{M \left( 1 - \dfrac{\sqrt{\theta} }{\sqrt{a^{2} + \theta}}\right) + \dot{a}^{2}}, 
\end{equation}
\begin{equation}
K^{\tau -}_{\,\ \tau} = \left\lbrace \dfrac{\ddot{a}  - \dfrac{a^2 M\sqrt{\theta}}{2(a^{2}+ \theta)^{3/2}} }{\sqrt{M \left( 1 - \dfrac{\sqrt{\theta} }{\sqrt{a^{2} + \theta}}\right) + \dot{a}^{2}}}\right\rbrace.
\end{equation}
In the following, we will apply these equations to analyze the energy conditions for the stability of gravastar in the thin shell using the Lanczos equations.

\subsection{Lanczos equations: Surface tension}
Now, to determine the stability in the thin shell, we will use the Lanczos equations, which derive from Einstein's equations applied to the hypersurface that joins the mass space-times, and are given by~\cite{Lobo:2015lbc}
\begin{equation}
S^{i}_{\, j} = - \frac{1}{8 \pi} \left(k^{i}_{\, j} - \delta^{i}_{\, j} \,\ k^{k}_{\,\, k}\right), 
\end{equation}
where $S^{i}_{\, j}$ is the surface energy-momentum tensor in $\Sigma$. Now, from the Lanczos equation in (2+1)-dimensional spacetime, the surface tension energy-momentum tensor $S^{i}_{\,j} = diag( - \sigma, \mathscr{P})$ where $\sigma$ is the surface density and $\mathscr{P}$ is the surface pressure~\cite{Perry:1991qq}. These are given as follows:
\begin{equation}
\sigma = - \frac{K^{\psi }_{\,\ \psi}}{4 \pi } = - \dfrac{1}{4 \pi a} 
\left[ \sqrt{M - \Lambda a^{2} + \dot{a}^{2}} - \sqrt{M \left( 1 - \dfrac{\sqrt{\theta} }{\sqrt{a^{2} + \theta}}\right)+ \dot{a}^{2}} \, \right], 
\label{a1}
\end{equation}
\begin{equation}
\mathscr{P} =  \frac{K^{\tau }_{\,\ \tau} + K^{\psi }_{\,\ \psi} }{8 \pi } 
=\frac{1}{8 \pi a} \left[\frac{M +\dot{a}^{2} +  \ddot{a} - 2\Lambda a^{2}}{\sqrt{M - \Lambda a^{2} + \dot{a}^{2}}} -\dfrac{M \left( 1 - \dfrac{\sqrt{\theta} }{\sqrt{a^{2} + \theta}}\right) + \dot{a}^{2} + \ddot{a}  - \dfrac{ a^2 M \sqrt{\theta}}{2 (a^{2}+\theta)^{3/2}} }{\sqrt{M \left( 1 - \dfrac{\sqrt{\theta} }{\sqrt{a^{2} + \theta}}\right) + \dot{a}^{2}} }  \right]. 
\label{a2}
 \end{equation}

Using the equations (\ref{a1}) and (\ref{a2}), we have: 
\begin{equation}
\sigma + 2\mathscr{P}=  \frac{K^{\tau }_{\,\ \tau}}{4 \pi } = \dfrac{1}{4\pi a}\left[  \dfrac{\ddot{a} - \Lambda {a^2} }{\sqrt{M - \Lambda a^{2} +\dot{a}^{2}}}   -   \dfrac{\ddot{a}  - \dfrac{ a^2 M \sqrt{\theta}}{2 (a^{2}+\theta)^{3/2}} }{\sqrt{M \left( 1 - \dfrac{\sqrt{\theta} }{\sqrt{a^{2} + \theta}}\right) + \dot{a}^{2}}} \right].					
\end{equation}
For the sake of discussion, let us consider a static solution where 
$a_{0} \in (r_{-},r_{+})$. So, we have:
\begin{equation}
{\sigma}(a_0) = -\frac{1}{4 \pi a_0} \left[  \sqrt{M - \Lambda a^2_0} - \sqrt{M \left( 1 - \dfrac{\sqrt{\theta}}{\sqrt{a^2_0 + \theta}}\right) } \, \right], 
\label{a04}
\end{equation}

\begin{equation}
{\mathscr{P}}(a_0) = \frac{1}{8 \pi a_0} \left[\frac{M - 2\Lambda a^2_0 }{\sqrt{M - \Lambda a^2_0  }} 
-\dfrac{M \left( 1 - \dfrac{\sqrt{\theta}}{\sqrt{a^2_0 + \theta}}\right)  - \dfrac{ a_0^2 M \sqrt{\theta} }{ 2(a^2_0 +\theta)^{3/2}} }{\sqrt{M \left( 1 - \dfrac{\sqrt{\theta}}{\sqrt{a^2_0 + \theta}}\right) } } \right],
\label{a05}  
\end{equation}

\begin{equation}
\sigma(a_0) + 2 {\mathscr{P}}(a_0) = \dfrac{1}{4\pi a_0}\left[\dfrac{ - \Lambda a^2_0 }{\sqrt{M  - \Lambda a^2_0 }}  
 + \dfrac{  \dfrac{ a_0^2 M \sqrt{\theta} }{ 2(a^2_0 +\theta)^{3/2}} }
 {\sqrt{M \left( 1 - \dfrac{\sqrt{\theta} }{\sqrt{a^{2}_0 + \theta}}\right)}}\, \right].	
\label{a06}					
\end{equation}

Now we can write the above equations in terms of the dimensionless parameters $\Tilde{\Lambda} = \Lambda a^2_0$ and $\Theta ={\sqrt{\theta}}/{a_{0}}$ as follows:
\begin{equation}
\Tilde{\sigma} = -\frac{1}{4 \pi} \left[  \sqrt{M - \Tilde{\Lambda}} - \sqrt{M \left( 1 - \frac{\Theta}{\sqrt{1 + \Theta^{2}}}\right) } \, \right], 
\label{a4}
\end{equation}
\begin{equation}
\Tilde{\mathscr{P}} =   \frac{1}{8 \pi} \left[ \frac{M - 2\Tilde{\Lambda} } {\sqrt{M - {\Tilde{\Lambda}}  }} 
-\dfrac{M \left( 1 - \dfrac{\Theta }{\sqrt{1 + \Theta^{2}}}\right)  - \dfrac{ \Theta M }{ 2(1 +\Theta^{2})^{3/2}}   }{\sqrt{M \left( 1 - \dfrac{\Theta }{\sqrt{1 + \Theta^{2}}}\right) } } \right],
\label{a5}  
\end{equation}
\begin{equation}
\Tilde{\sigma} + \Tilde{{\mathscr{P}}} = \frac{1}{8\pi}\left[\frac{- M  }{\sqrt{M  - {\Tilde{\Lambda}} }}  + \dfrac{M \left( 1 - \dfrac{\Theta }{\sqrt{1 + \Theta^{2}}}\right)+ \dfrac{ \Theta M }{2\left( 1 +\Theta^{2} \right)^{3/2} } }
 {\sqrt{M \left( 1 - \dfrac{\Theta }{\sqrt{1 + \Theta^{2}}}\right) }}\, \right],	
\label{aa6}					
\end{equation}
\begin{equation}
\Tilde{\sigma} + 2 \Tilde{{\mathscr{P}}} = \frac{1}{4\pi}\left[\frac{ - {\Tilde{\Lambda}} }
{\sqrt{M  - {\Tilde{\Lambda}} }}  + \dfrac{ \dfrac{ \Theta M }{2\left( 1 +\Theta^{2} \right)^{3/2} } }
 {\sqrt{M \left( 1 - \dfrac{\Theta }{\sqrt{1 + \Theta^{2}}}\right) }}\, \right],	
\label{a6}					
\end{equation}
where we have defined $\Tilde{\sigma}= a_0\sigma(a_0) $ and $\Tilde{\mathscr{P}}= a_0\mathscr{P}(a_0)$.
Note that the energy density $\Tilde{\sigma}$ is negative, but the pressure $\Tilde{\mathscr{P}}$ is positive. Furthermore, in this infinitely thin shell, the radial pressure is zero. It can be noticed that in both states $\Tilde{\sigma} +  \Tilde{\mathscr{P}}$ and $\Tilde{\sigma} + 2 \Tilde{{\mathscr{P}}}$ are positive, characteristic of the transition between the thin shell 
and the outer region~\cite{Rahaman:2012wc}. 
It is also interesting to observe that even when the cosmological constant is zero ($\Lambda=0$) or $l\rightarrow\infty$, we still have the stability condition satisfied with $\Tilde{\sigma}<0$ and $\Tilde{\mathscr{P}}>0$ due to the nocommutativity effect. 
To avoid issues with units, we also solved our equations in dimensionless form.
Thus, for $\Tilde{\Lambda}=0$, equations \eqref{a4}, \eqref{a5} and \eqref{a6} are respectively given by
\begin{equation}
\Tilde{\sigma}= -\frac{1}{4 \pi} \left[ \sqrt{M }- \sqrt{M \left( 1 - \dfrac{\Theta }{\sqrt{1 + \Theta^{2}}}\right)} \,\ \right], 
\label{a7}
\end{equation}
\begin{equation}
\Tilde{\mathscr{P}}=  \frac{1}{8 \pi } \left[ \frac{M}{\sqrt{M}} 
-\dfrac{M \left( 1 - \dfrac{\Theta }{\sqrt{1 + \Theta^{2}}}\right)  - \dfrac{ \Theta M }{2(1 +\Theta^{2})^{3/2}}  }{\sqrt{M \left( 1 - \dfrac{\Theta }{\sqrt{1 + \Theta^{2}}}\right)} } \right],
\label{a8}  
\end{equation}
\begin{equation}
\Tilde{\sigma} + \Tilde{{\mathscr{P}}} = \frac{1}{8\pi}\left[\frac{- M  }{\sqrt{M }}  + \dfrac{M \left( 1 - \dfrac{\Theta }{\sqrt{1 + \Theta^{2}}}\right)+ \dfrac{ \Theta M }{2\left( 1 +\Theta^{2} \right)^{3/2} } }
 {\sqrt{M \left( 1 - \dfrac{\Theta }{\sqrt{1 + \Theta^{2}}}\right) }}\, \right],	
\label{ab6}					
\end{equation}
\begin{equation}
\Tilde{\sigma} + 2 \Tilde{{\mathscr{P}}}= \frac{1}{4\pi}\left[  
  \frac{ \dfrac{ \Theta M }{2\left( 1 +\Theta^{2} \right)^{3/2} }}
{\sqrt{M \left( 1 - \dfrac{\Theta }{\sqrt{1 + \Theta^{2}}}\right) }}\,\right] .	
\label{a9}					
\end{equation}

Now, for $\Theta\ll 1\, (\theta\ll 1)$ we find
\begin{equation}
{\sigma}\approx -\frac{\sqrt{M}\Theta}{8 \pi a_0}=-\frac{\sqrt{M\theta}}{8 \pi a^2_0}, 
\label{a10}
\end{equation}
\begin{equation}
{\mathscr{P}}\approx \dfrac{\sqrt{M}\Theta}{8 \pi a_0}=\frac{\sqrt{M\theta}}{8 \pi a^2_0},
\label{a11}  
\end{equation}
\begin{equation}
\sigma + 2 {\mathscr{P}}\approx \frac{\sqrt{M}\Theta}{8\pi a_0}=\frac{\sqrt{M\theta}}{8\pi a_0^2}.	
\label{a12}					
\end{equation}
However, from the above equations, we have the following equation of state
\begin{eqnarray}
 {\sigma} + {\mathscr{P}} =0,\, \quad\quad 
 {\mathscr{P}}=-{\sigma}=\rho.
\end{eqnarray}
{Thus, since $\sigma<0$, we have $p=\rho$ in the thin shell, an effect due to noncommutativity.
On the other hand, considering $\Lambda=0$, inside the shell, we have the state equation $p=-\rho$, 
with $\rho=\sigma\sim \sqrt{\theta}/M_0r^3>0$.
Therefore, the equation of state, $p=-\rho$, represents a repulsive pressure. 
In the context of an accelerated Universe, this would be related to dark energy arising due to the effect of noncommutativity
(\textit{$\theta$-dark energy}). Here, we show that by setting $\Lambda=0$, the noncommutativity parameter $\theta$ plays the role of the cosmological constant for gravastar formation and stability.}

On the other hand, by considering $\Tilde{\Lambda}$ too large, we can write the equations for $\Tilde{\sigma}$ and $\Tilde{\mathscr{P}}$ as follows:
\begin{equation}
{\sigma} \approx -\frac{\sqrt{-\Tilde{\Lambda}}}{4 \pi a_0}=-\frac{\sqrt{-{\Lambda a^2_0}}}{4 \pi a_0}, 
\label{alambg}
\end{equation}
\begin{equation}
{\mathscr{P}} \approx \frac{\sqrt{-\Tilde{\Lambda}}}{4 \pi a_0}=\frac{\sqrt{-{\Lambda a^2_0}}}{4 \pi a_0}.
\label{plambg}  
\end{equation}
For this case, with ${\Lambda}<0$, we obtain the following equation of state
\begin{eqnarray}
{\mathscr{P}}=-{\sigma}.
\end{eqnarray}
By comparing the results above, we find a relationship between ${\Lambda}$ and $\theta$ given by
\begin{eqnarray}
 \sqrt{-{\Lambda}a^2_0} = \dfrac{\sqrt{M}\Theta}{2}=\dfrac{\sqrt{-M_0\theta}}{2a_{0}}, \qquad \mbox{or} \qquad 
 \theta = \dfrac{4  a^{4}_{0} \Lambda}{M_{0}}.
\end{eqnarray}
From the result above, we can also write a relationship between $\Theta$ and $\Tilde{\Lambda}$, that is, 
$\Theta=({4\Tilde{\Lambda}/M_0})^{1/2}$ (which can also be obtained from equation (\ref{aa6}) for $\Tilde{\sigma} + \Tilde{{\mathscr{P}}}=0$, considering $\Tilde{\Lambda}$ and $\Theta$ small).
Now admitting $M_{0} = {M_{BH}}/{M_{\odot}}$, where $M_{BH} $ is the mass of the black hole and 
$M_{\odot}=1.989 \times 10^{30}$ kg is the solar mass,
then, for $M_{BH}=10M_{\odot}$, $a_0\approx 29.5 \times 10^{3}$ m (radius of the black hole) and cosmological constant $\Lambda=1.088 \times 10 ^{-58}$ m$^{-2}$, we obtain the following value for the parameter $\theta$:
\begin{eqnarray}
 \theta \approx 3.296 \times 10^{-41} m^{2}=\left[3.4371 \times 10^{4} GeV\right]^{-2}
=\left[3.4371\times 10\, TeV \right]^{-2} .
\end{eqnarray}
Therefore, we found a value of $\theta\sim [10$ TeV$]^{-2}$ or $ \sqrt{\theta}\sim [10$ TeV$]^{-1}$, and an energy scale $\Lambda_{NC}=1/\sqrt{\theta}\sim 10$ TeV in accordance with results obtained 
in the literature~\cite{Mocioiu:2000ip,Chaichian:2000si,Chaichian:2002ew,Hinchliffe:2002km,Falomir:2002ih} 
(see also Ref.~\cite{Rivelles:2011gq} for other limits of $\theta$ and~\cite{Vagnozzi:2022moj} using the Event Horizon Telescope (EHT) observations of Sagittarius A$^{\ast}$).

Here it is opportune to mention that noncommutativity plays a vital role in black hole physics. Some effects that disappear in the usual case can be observed due to noncommutativity. 
For example, in~\cite{Anacleto:2012du}, it was found that when the circulation parameter is zero, the differential cross section goes to zero, and thus there is no analogous Aharonov-Bohm effect. On the other hand, due to noncommutativity, the analogous Aharonov-Bohm effect persists even when the circulation parameter is set to zero. 
By considering the noncommutative BTZ black hole, Anacleto and collaborators~\cite{Anacleto:2014cga} showed that due to the noncommutativity, the gravitational Aharonov-Bohm effect is observed when the circulation parameter goes to zero. 
Furthermore, in~\cite{Anacleto:2019tdj,Anacleto:2022shk,Campos:2021sff}  the noncommutativity effect was also explored in calculating the differential cross section, absorption, quasinormal modes and shadow radius and verified that these quantities are proportional to the noncommutativity parameter when the mass goes to zero. 
In addition, the stability condition and remainders for the noncommutative BTZ black hole and the noncommutative Schwarzschild black hole via calculating specific heat were examined in~\cite{Anacleto:2020efy,Anacleto:2020zfh}.

\begin{figure}[!htb]
\centering
\includegraphics[scale=0.65]{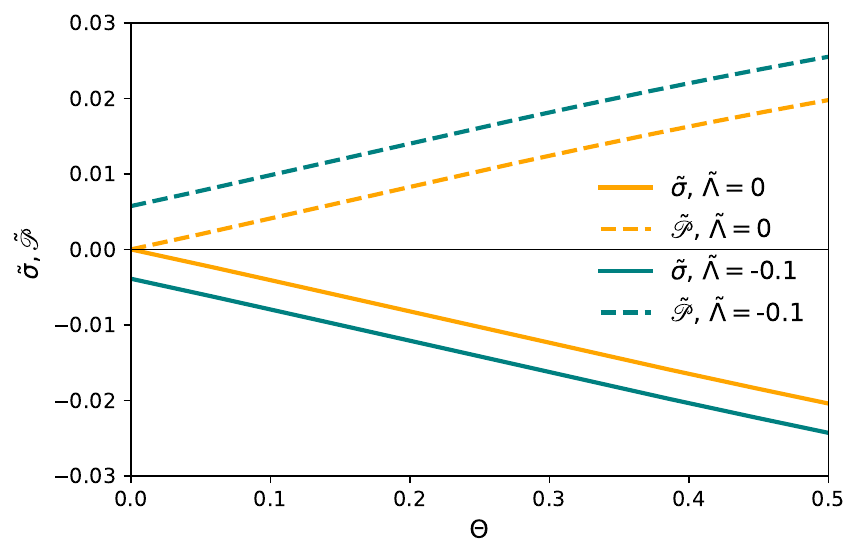}
\captionsetup{justification=raggedright,singlelinecheck=false}
\caption{\footnotesize{Energy density $\Tilde{\sigma}$ (solid line) and pressure $\tilde{\mathscr{P}}$ (dashed line) as a function of parameter $\Theta$ for $\Tilde{\Lambda} = 0$ (orange) and $\Tilde{\Lambda} = -0.1$ (teal), assuming $ M=1 $.}}
\label{EC1}
\end{figure}

\begin{figure}[!htb]	
\centering 
\includegraphics[scale=0.65]{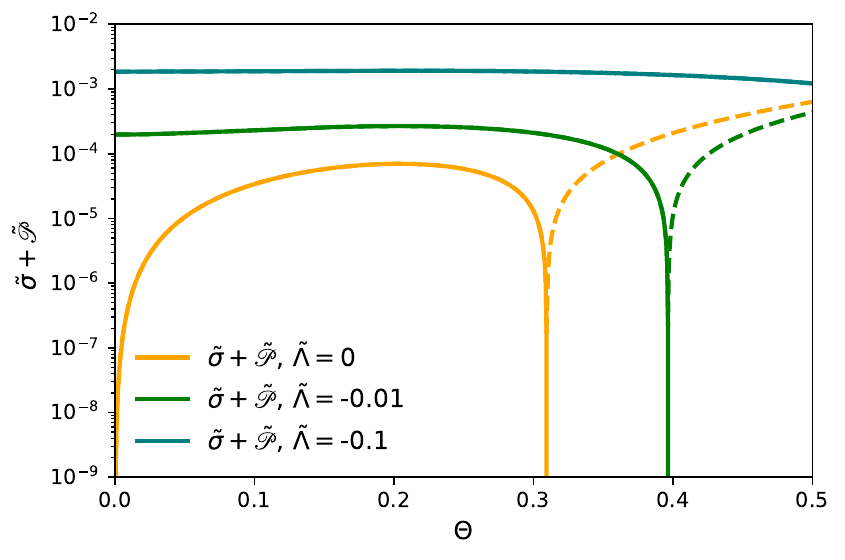}
\captionsetup{justification=raggedright,singlelinecheck=false}
\caption{\footnotesize{$\tilde{\sigma}+\tilde{\mathscr{P}}$  as a function of $\Theta$ for  $\Tilde{\Lambda}=0$ (orange), $\Tilde{\Lambda}=-0.01$ (green), and $\Tilde{\Lambda}=-0.1$ (teal), assuming $M=1$. Solid lines shows the positive value of $\tilde{\sigma}+\tilde{\mathscr{P}}$, dashed lines show the negative values, or $-(\tilde{\sigma}+\tilde{\mathscr{P}})$. }}
\label{EC2}
\end{figure}

\begin{figure}[!htb]	
\centering 
\includegraphics[scale=0.65]{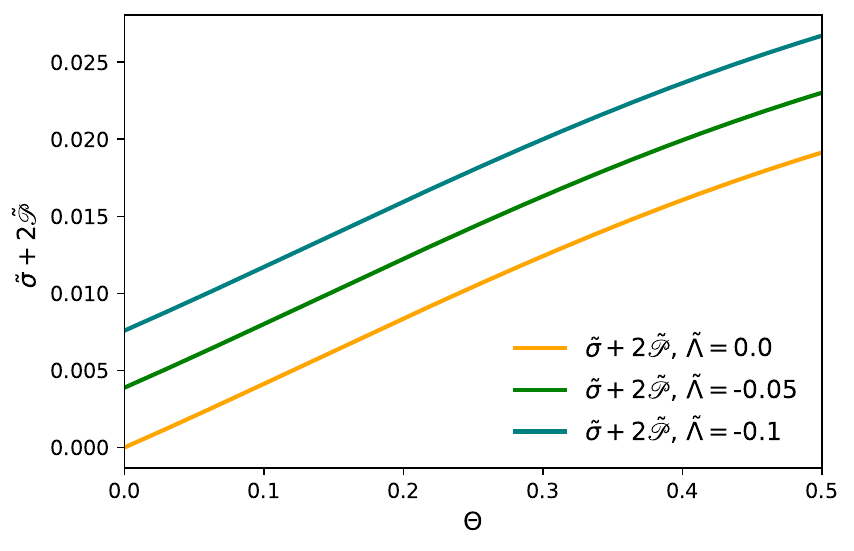}
\captionsetup{justification=raggedright,singlelinecheck=false}
\caption{\footnotesize{$\tilde{\sigma}+2\tilde{\mathscr{P}}$ as a function of $\Theta$ for $\Tilde{\Lambda}=0$ (orange),  $\Tilde{\Lambda}=-0.05$ (green) and $\Tilde{\Lambda}=-0.1$ (teal), assuming $M=1$.}}
\label{EC3}
\end{figure}

In Fig.~\ref{EC1}, we use the equations (\ref{a4}) and (\ref{a5}) to show the energy density and the pressure as a function of the noncommutativity parameter $\Theta$ for $\Tilde{\Lambda} < 0$ and  $\Tilde{\Lambda} = 0$.  We note that the energy density is negative and the pressure is positive for both cases. 
In Fig.~\ref{EC2} and Fig.~\ref{EC3}, we employ the equations (\ref{a4}),~(\ref{a5}), and (\ref{a6}) to show the suitable quantities $\tilde{\sigma}+\tilde{\mathscr{P}}$ and $\tilde{\sigma}+2\tilde{\mathscr{P}}$ respectively, that allow us to complete the evaluation of the energy conditions for gravastar stability as a function of $\Theta$ for $\Tilde{\Lambda} = 0$ and $\Tilde{\Lambda} < 0$.

The energy conditions require that, if $\sigma \geq 0$ and $\sigma + \mathscr{P} \geq 0 $ are satisfied, then the weak energy condition (WEC) holds. We have, by continuity, the null energy condition (NEC) valid, because  $\sigma + \mathscr{P} \geq 0 $. For the strong energy condition (SEC) to be proven, it is required that $\sigma + \mathscr{P} \geq 0$ and $\sigma + 2 \mathscr{P} \geq 0$~\cite{Banerjee:2013qja,Horvat:2007qa}. In our calculations, we show that $\sigma$ is negative, however, the pressure $ \mathscr{P}$ is positive (Fig. \ref{EC1}). 
The positive pressure contributes to maintain $\tilde{\sigma}+\tilde{\mathscr{P}} \geq 0$ for sufficiently small $\Theta$ (Fig. \ref{EC2}),
as well as $\tilde{\sigma}+2\tilde{\mathscr{P}} \geq 0$ (Fig. \ref{EC3}).
Therefore, the shell contains matter that violates only the  weak energy condition (WEC) and obeys the null and strong energy  conditions when $\Theta$ is small enough.

\section{Conclusion}
\label{Conc}
In this paper, we have considered the noncommutative BTZ black hole with noncommutativity introduced via the Lorentzian distribution~%
Therefore, we have performed its thermodynamic analysis and investigated a thin-shell gravastar model.
We show that noncommutativity plays the role of regularizing the temperature of the three-dimensional Schwarzschild anti-de Sitter black hole. 
In addition, by computing entropy we have found a logarithmic correction term, $2\pi\sqrt{\theta}\ln r_h$. 
Furthermore, we examine the stability of the black hole by calculating the specific heat capacity.
We have found that for a given minimum radius dependent on the parameter $\theta$ the specific heat capacity goes to zero, indicating the formation of a hole remnant as the final stage. 
By analyzing the gravastar model of a thin-shell noncommutative BTZ black hole, 
we have found that the energy density $ \sigma $ is negative and the pressure $ \mathscr{P} $ is positive. 
Also, we have verified that the state $\Tilde{\sigma} + \Tilde{\mathscr{P}}$ and $\Tilde{\sigma} + 2\Tilde{\mathscr{P}}$ are positive for quite small $\Theta$ values.
Moreover, even for ${\Lambda}=0$, the results obtained above are maintained. 
{Besides, even in the absence of the cosmological constant, we have obtained that the condition $p=\rho$ in the thin shell is satisfied with $\rho\approx{\sqrt{M\theta}}/{(8 \pi a^2_0)}$ (for small $\theta$) 
and in the inner region the equation of state $p= -\rho$ ($\rho\sim \sqrt{\theta}/M_0r^3 $) is also satisfied showing the effect of noncommutativity for repulsive pressure and dark energy.}
Therefore, in our calculations we have shown that the $\theta$ parameter plays the role of the cosmological constant for the gravastar energy stability condition. 
Hence, we have found a relationship between the noncommutativity parameter and the cosmological constant. 
Thus, we have estimated a value of the order of $\theta\sim [10$ TeV$]^{-2}$ for the noncommutativity parameter. 
{By comparing with Ref.~\cite{Banerjee:2016bry} which deals with the gravastar solution in noncommutative geometry. The authors introduced noncommutativity via Gaussian distribution and verified that $\sigma<0$ and $p>0$ in the thin shell. In our analysis, we have introduced noncommutativity using a Lorentzian distribution that made it possible to find the conditions on the energy density and pressure in the thin shell in the limit of small $\theta$, as well as for the zero cosmological constant. Furthermore, it was possible to obtain a bound for the noncommutativity parameter.}

\acknowledgments

We would like to thank F. A. Brito for useful discussions and CNPq, CAPES, CNPq/PRONEX/FAPESQ-PB  (Grant nos. 165/2018 and 015/2019), for partial financial support. MAA acknowledge support from CNPq (Grant nos. 306398/2021-4).

\end{document}